\documentclass[10pt,a4paper]{article}
\usepackage{amsmath,amssymb}
\usepackage{epsfig,graphicx}

\begin{document}

\textwidth=135mm

\textheight=200mm

\begin{center}

{\bfseries 
The Zoo of Neutron Stars}

\vskip 5mm

S.~B.~Popov$^\dag$\footnote{{\bf e-mail: polar@sai.msu.ru}}

\vskip 5mm

\small{\it $^\dag$ Sternberg Astronomical Institute,
Universitetski pr. 13, 119992 Moscow, Russia}\\
\small{\it Cariplo Foundation Fellow}



\end{center}

\vskip 5mm

\centerline{\bf Abstract} 

In these lecture notes I briefly discuss the present day situation and new
discoveries in astrophysics of neutron stars focusing on isolated objects.
The latter include soft gamma repeaters, anomalous X-ray pulsars, central
compact objects in supernova remnants, the Magnificent seven, and rotating
radio transients. In the last part of the paper I describe available tests
of cooling curves of neutron stars and discuss different additional
constraints which can help to confront theoretical calculations of cooling
with observational data.

\vskip 10mm

\section{Introduction}

 As the Moscow Zoo, the Zoo of neutron stars (NSs) can be separated into {\it old} and {\it new} parts.
The {\it old} part includes classical radio pulsars and accreting NSs in close binary systems.
This territory started to be filled with "animals" already in 60s, and most of the "beasts" are well known
even to general public. 
The {\it new} one is mostly populated by isolated NSs which belong to five main types which have been mainly recognized in the last 10 years or so. These five species are: soft gamma repeaters (SGRs), anomalous X-ray pulsars (AXPs), central compact objects in supernova remnants (CCOs in SNRs), the Magnificent seven (M7), and rotating radio transients (RRATs).
May be in near future more types will be recognized (for example, related to
unidentified EGRET sources, in this respect data from the GLAST mission will
be of crucial importance).
In the following section I try to give an extremely brief guide for this {\it new} territory of the Zoo of NSs.

As due to a strict page limit only few words can be said about each type,
at first I'd like to give a list of reviews on each of the types mentioned below. Of
course, I even cannot list all reviews on the subjects, so the choice is
very subjective, but, still, representative.

As a general introduction to the Zoo of NSs one can take the short encyclopedic article by Baym and Lamb \cite{bl2005} and references therein.
SGRs and AXPs are very well described in \cite{wt2004}. Theory of magnetars was reviewed many times, 
one can use, for example, the review by Heyl \cite{h2005}.
A perfect recent review on AXPs can be found in \cite{kaspi2006}.
Observations of SGRs are also reviewed in \cite{met2006}.
To have an impression of how CCOs look like, one can take the brief paper \cite{pst2004}. 
A huge set of Chandra results on observations of SNRs (including CCOs) can be found in \cite{w2005}.
An extensive search for compact sources in SNRs was presented in \cite{k2004}.
The Magnificent seven attracted much interest in last few years. Two interesting reviewing papers were published recently by Tr\"umper \cite{t2005} and Haberl \cite{h2006}. RRATs appeared in the Zoo very recently, so there are no reviews, yet. One should refer to the original paper \cite{mac2006}. 

In the last part of this note I speak about tests of theories of the thermal
evolution of NSs. A very good recent review on the cooling can be found in
\cite{pgw2005}.

All subjects touched in this paper have been excellently reviewed during the
conference "Isolated Neutron Stars: from the Interior to the Surface"
(London, April 2006). Proceedings of the meeting will be published soon in
the journal Astrophysics and Space Science, and this volume is going to be
the best set of materials on the subject in the near future.\footnote{Many
presentations from this conference are available on the Web:
http://www.mssl.ucl.ac.uk/\%7Esz/Conference\_files/index.html.}

Finally, a russian-speaking reader can have a look at our review \cite{pp2003}, where all subjects mentioned here are discussed in more details.\footnote{This review is available on the Web at this URL:
http://xray.sai.msu.ru/$\sim$polar/html/kniga.html.}

In the Table 1 I give the list of sources. Mostly, data on SGRs  are taken from \cite{wt2004}, on AXPs -- from \cite{kaspi2006},
on CCOs -- from \cite{pst2004}, on the M7 -- from \cite{h2006}. However, some additions from other publications are made.
In particular, I want to underline a recent determination of spin period of
RX J1856-3754 \cite{mt2006}.

\section{Soft gamma repeaters}

The first burst of a SGR was detected long ago on March 5, 1979. The source
was recognized as an object with spin period about 8 seconds in the SNR N49
in the Large Magellanic Cloud \cite{m1979, v1979}. Since then three more
sources of this type have been discovered, and a few candidates are known,
too.

Spin periods in the range $\sim 5$~--~8~s and associations of some of SGRs with young SNRs undoubtly point towards young NSs. Before December 2004 three main types of burst have been recognized: weak, intermediate, and giant. Weak bursts are most numerous, hundreds of them have been detected. Their typical durations are about $0.1$~s, and luminosities are $<10^{41}$~erg~s$^{-1}$. They usualy have single-peak structure, and tend to concentrate around periods of activity, during which more rare and energetic bursts appear. Intermediate bursts show a variety of morphologies. Their luminosities are about $10^{41}$~--$10^{43}$~erg~s$^{-1}$.  More energetic events are classified as giant bursts.
The historic March 5 burst of SGR 0526-66 is one of them. A similar event was detected from SGR 1900+14 on August 27, 1998. Some researchers classify the June 18, 1998 burst of SGR 1627-41  as a giant one \cite{m2005}, but it did not have a typical pulsating ``tail'', so this classification is usualy doubted.

The last giant burst of SGR 1806-20, observed on December 27, 2004, is often  marked out, and classified as a {\it hyper flare}. The reason is simple: its energetics is about 2 orders of magnitude higher than in the case of other giant bursts.

One of the main recent discovery made in observations of these sources is the detection of quasiperiodic oscillations during giant flares \cite{i2005, ws2005}.   Results were obtained for bursts of SGR 1900+14 and SGR 1806-20. Oscillation frequencies are about tens of Hz. Most probably, they reflect torsional vibrations of the neutron star crust. 

Another intriguing discovery is related to observations of the GRB 051103  \cite{f2006}.
The authors provide evidence that this short gamma-ray burst has not the cosmological origin, but is a hyper flare of a SGR in the group of galaxies around M81. If confirmed, it is the first clear observation of a SGR outside the local group of galaxies (see, however, \cite{p2005,ps2006}).

Standard interpretation of SGRs is related to {\it magnetars} -- NSs releasing their magnetic energy.
The reasons is that neither rotational, nor thermal, or any other kind of energy stored in a NS can explain the observed phenomenae. The same interpretation is applied to AXPs, which are assumed to be at least cousins of SGRs.

\section{Anomalous X-ray pulsars}

AXPs have been recognized as a separate class among X-ray pulsars in 1995 \cite{s1995, p1995}.
Their periods are clustered in a narrow range between $\sim 5$ and 12 s, they continuously spin--down, 
their luminosities are stable and somehow smaller than for other X-ray pulsars, and, finally, no binary companions have been found for them. 
Now 9 objects of this type are known. Some of them are situated inside SNRs.

Connection between AXPs and SGRs is supported by the following arguments.
The first one is the most obvious: they have similar $P$ and $\dot P$ values.
Then, astrophysical manifestations of both types can be quite similar.
In quiescent state SGRs share similar properties with AXPs. For example, SGR 0526-66 shows no signs of bursts since early 80s, and it looks like a typical AXP. On the other hand, AXPs can produce bursts
\cite{g2003}, which are very similar to weak bursts of SGRs. Finally, for most of AXPs their thermal or rotational energies are not sufficient to explain the observed activity, as it is in the case of SGRs.

One of the main recent result in the field, in my opinion,  is the discovery of a remnant disc around AXP 4U 0142+61 \cite{w2006}. 
The possibility of the existence of remnant discs formed due to fall-back of matter after a SN explosion was discussed since long ago. The idea of active discs of this kind was considered as the main alternative to the magnetar scenario (at least in the case of AXPs) \cite{a2000}. Such discs could contribute to the spin down of NSs, and, probably, to its luminosity. However, the discovered disc is most probably the passive one. I.e., it has nothing to do with the present day activity of the AXP.\footnote{See, for example, \cite{kaspi2006} for the critics of the debris disc models.}
 
Another result which necessarily should be mentioned, is the discovery of hard tails in spectra of several AXPs \cite{mol2004, rev2004,dh2006}. This was done thanks to observations onboard INTEGRAL satellite.
Good sensitivity of this observatory in hard X-ray range resulted in detection of emission above 10 keV (up to 150 keV) from AXPs 1RXS J1708-4009, 4U 0142+61, 1E 1841-45,  and 1E 2259+586.
This result poses new questions in front of the theory of magnetar emission.

Finally, it is important to note that AXPs (and, probably, SGRs and M7) should not be considered absolutely radio-silent any more. 
Radio emission was detected on VLA from the transient AXP XTE J1810-197 \cite{hg2005}. Earlier,
detection of low frequency radio signals from an AXP and from a member of the M7  have been reported by the Pushchino group \cite{m2005}.

\section{Central compact objects in supernova remnants}

 CCOs are defined as X-ray sources with thermal-like spectra observed close to centers of non-plerionic SNRs without any counterparts in radio and gamma wavebands. 
They show blackbody temperatures about few hundreds of eV, and have luminosities $\sim 10^{33}$~--~$10^{34}$~erg~s$^{-1}$.
About 10 of such sources are known, including famous RCW103, Cas A, Pup A, and Kes 79.
In different papers one can find slightly different lists of sources depending on the criteria used to select them. Also, the number is continuosly increasing. For example, the announcement of the last candidate
discovery appeared during the preparation of this manuscript \cite{rb2006}.

Some of the sources (Cas A, Vela Junior) have according to spectral fits surprizingly small emitting areas.
Typical sizes for them are $<2$~km, well below the size a normal NS. This puzzle remains unsolved.

Recently, a clear 6.7 hour period was confirmed in RCW103 \cite{luca2006}.
The origin of the periods is unclear: it can be an orbital or a spin period.
In the first case, there are strict limits on the secondary companion: it
can not be a normal star with $M>0.4\, M_{\odot}$. The companion can be, for
example, another NSs \cite{p1998}. If the newborn NS which produced the SNR
has a remnant disc around it, then the older NS can accrete from that disc
when passing close to the companion. Another possibility is that the
secondary companion is a very low-mass star inside the magnetosphere of a NS
(a system similar to so-called {\it polars}). However, such a system can
hardly be formed without a significant kick velocity, but the proximity of
the source to the geometrical center of the SNR point to low velocity
$<150$~km~s$^{-1}$. De Luca et al. \cite{luca2006} favor the idea that the
observed period is the spin period of an extreme magnetar with $B >
10^{15}$~G. Even with such field a NS cannot spin down to 6.7 hours via
magneto-dipole (or longitudinal current) losses during the lifetime of the
remnant. So, a kind of propeller mechanism should be working. Appearence of
such a field is doubtful, in my opinion, as in the case of RCW 103 it has to
be the fossil field not significantly amplificated due to some kind of a
dynamo mechanism, as there are no traces of huge energy input into the SNR
(see \cite{v2006} for discussion of such limitation). Anyway, whatever is
the real nature of the source it is very peculiar and puzzling.

Another interesting result is related to the source Puppis A. Winkler and Petre \cite{wp2006} found that this object has one of the largest spatial velocity among all known NSs, $\sim 1500$~km~s$^{-1}$. This is the first case when proper motion of a CCO is measured directly. 

\section{The Magnificent seven}

The first source of this type, RX J1856-3754, was discovered 10 years ago
\cite{w1996} in the ROSAT data. Later on six other similar objects were
identified, also in the data obtained with the ROSAT. All seven are
recognized to be relatively close-by (less than few hundred pc), middle-age
(about several hundred thousands years) isolated NS emitting soft X-rays due
to cooling. The latter is confirmed by the blackbody shapes of their
spectra. Typical temperatures are about 50~--~100~eV. At least five out of
the seven show spin periods in the range 3-12~s. Recently, discovery of
pulsations have been reported also for RX J1856.5-3754 \cite{mt2006} and RX
J1650.3+3249 (see \cite{h2006} and references therein).
The case of RX
J1856-3754 is the most spectacular as before very recent time
only very strict limits on any kind of pulsations have been reported
\cite{b2003}. 
Some of the Seven have very weak optical counterparts. For the
brightest one (RX J1856-3754) the trigonometric parallax and proper motion
are known \cite{k2002}. These data provide a possibility to reconstruct 3D
trajectory, and so to identify the birth site of the NS.

Population synthesis studies \cite{p2003} show that the M7 are related to
the Gould Belt -- local structure with the age $\sim 30$~--~50~Myrs formed
by massive stars. Reconstruction of trajectories of NSs confirmed this
conclusion.
In the solar vicinity this NSs outnumber radio puslars of
the same age. This means that the M7-like objects can be one of the most
typical young NSs with galactic birth rate larger than that of normal radio
pulsars.

XMM-Newton observations made possible to detect wide absorption features in spectra of several among the M7. The origin of these features is not known (see \cite{h2006} for references and more detail description of the results presented next). They can be proton (or ion) cyclotron lines in strong ($>10^{13}$~G) magnetic field, or absorption lines due to atomic transitions. For two of the M7 (RBS 1223 and RX J0720.4-3125) spectra are shown to be phase dependent. In the case of RX J0720.4-3125 the X-ray spectrum and pulse profile are changing with time with a possible period about 7 years, which is attributed to the free precession of the NS.
The Seven objects seem to be the best laboratory to study NS atmospheres and, probably, internal structure \cite{t2005}.

 Probably, the M7 are not absolutely unactive in the radio band.
A discovery of a radio signal from 1RXS J214303.7+065419+06 was recently reported in \cite{m2006}. Still the result has not very high significance and has to be verified.

 Unfortunately, up to now only seven objects of this type are known. The
last one was identified already in 2001 \cite{zam2001}. However, population
synthesis studies predict that up to several dozens of sources in the ROSAT
catalogue are waiting for their identification. Our recent calculations
(\cite{pos2006} and Popov et al., work in progress) demonstrate that new
candidates with count rates 0.1-0.01 ROSAT counts per second should be
younger and hotter than the known seven sources, and should originate from
rich OB associations behind the Gould Belt. If not identified in the ROSAT
data they will we uncovered by eRosita detector on-board future sattelite
SRG. 

\begin{table}
\begin{tabular}{l|c|c|c}
\hline
&&& \\
Name & $P$, s & $\dot P/10^{-13}$ & Comments\\
\hline
\hline
&&& \\
{\bf SGRs} & & & {\bf Giant flares} \\
0526-66 &8.0&660&5  March 1979 \\
1627-41 &6.4&--&18 June 1998 (?) \\
1806-20  &7.5&830--4700&27 Dec 2004 \\
1900+14 &5.2&610--2000&27 Aug 1998 \\
\hline
&&& \\
{\bf AXPs} &&&  {\bf Remarks}\\
CXO 010043.1-72  &8.0&190& SMC\\
4U 0142+61         &8.7&20& Remnant disc\\
1E 1048.1-5937    &6.4&270& Bursts\\
CXOU J164710.2-455216 & 10.6 & -- & Westerlund 1\\
1 RXS J170849-40 &11.0&190& \\
XTE J1810-197    &5.5&50& Transient, bursts\\
1E 1841-045       &11.8&420& SNR Kes 73\\
AX J1845-0258    &7.0&--& Transient, SNR G29.6+0.1 \\
1E 2259+586      &7.0&4.8& Bursts, SNR CTB109\\
\hline
&&&\\
{\bf CCOs} &&& {\bf SNR}\\
J000256+62465 &--&& G117.9+0.6 \\
J082157.5-430017 &--&&Pup A \\
J085201.4-461753 &--&& G266.1-1.2 \\
J121000.8-522628 &0.424& 0.13& G296.5+10.0 \\
J161736.3-510225 &6.7 hours&& RCW 103 \\
J171328.4-394955 &--&& G347.3-0.5 \\
J181852.0-150213 & -- && G15.9+0.2 \\
J185238.6+004020 &0.105&& Kes 79 \\
J232327.9+584843 &--&& Cas A \\
\hline
&&& \\
{\bf M7} &&& {\bf Optical magnitude}\\
RX J0420.0-5022 &3.45&$<92$& B = 26.6\\
RX J0720.4-3125 &8.39&0.698& B=26.6 \\
RX J0806.4-4123 &11.37&$<18$& B$>$24\\
RBS 1223 &10.31&1.120&  m$_{50ccd}$=28.6\\
RX J1650.3+3249 &6.88 (?)&--& B=27.2\\
RX J1856.5-3754 &7.05&--& B=25.2\\
RBS 1774 &9.44&$<60$& B$>$26\\
\hline
\end{tabular}
\end{table}

\section{Rotating radio transients}

The latest major discovery in the field of isolated NSs was made just a year ago.
A new type of sources was discovered -- Rotating RAdio Transients (RRATs) \cite{mac2006}.
These objects emit very short bursts of radio waves. With a complicated analysis it became possible to measure periods, and for some sources even period derivatives. Periods are about 0.4-7 s, and $\dot P$ about $10^{-13}$ s~s$^{-1}$, so we can be more or less sure that the objects are NSs. 
No traces of binarity have been noticed. 
On the $P-\dot P$ diagram the sources are situated in the region of highly magnetized radio pulsars, close (but not very) to the region where SGRs and AXPs are found. Note, that the M7 occupy the same part of the diagram.

 Only 11 objects are known up to now. But, as the authors of the discovery estimated, their number and birth rate can be very high, even higher than that of normal radio pulsars.
If RRATs do not represent a completely new population of NSs then the only type which can compare by the birth rate are the M7 \cite{ppt2006}. One of RRATs have been detected in X-rays as a thermal source by Chandra \cite{r2006}. This makes the possible connection between the M7 and RRATs very plausible.

\section{Astronomy meets QCD: tests of cooling curves}

NSs are one of the most favorite astronomical objects in the physics community due to several reasons: strong gravity, strong magnetic fields, huge density inside, etc. The latter one is particularly interesting in respect with the subject of the school (see contributions by D. Blaschke, H. Grigorian, J. Berdermann, I. Parente, N. Ippolito). 
NSs are objects where astronomy meets QCD.

  Testing the behaviour of matter in different regions of the QCD phase
diagram is an extremely important, but difficult, task. The region
corresponding to  high density, but low temperature is not studied by
terrestrial laboratory experiments, yet. Observations of cooling NSs give an
opportunity to get indirect information about physical processes in this
region. The idea is to compare calculatated cooling curves with some data
obtained from astronomical observations.
In this section I discuss different approaches to do this.

The most standard test (I dub it below as the  T--t test) is the following. 
One just selects sources with known ages and temperatures and confront data
points with theoretical cooling curves.
Naively it is assumed that if all data points can be covered by cooling
tracks then the model can be considered to be in correspondence with
observations.

The main advantages of this test from the point of view of its 
use by the community are the following two:

\noindent
1. It is clear and direct.

\noindent
2. Everybody who calculate the theoretical curves can do the test as
observational data is available in the literature.

The test is widely used and was very well described many times (see, for
example, \cite{page2004} and references therein). So, we do not give many
details. Let us just specify few disadvantages which can be overcomed if one
uses additional tests and considerations.

A). Well determined temperatures and, especially, ages are known for very
few objects. So, statistics is not very large.

B). Usualy both -- temperature and age -- are known with some uncertainties
or depend on a chosen model.

C). Objects with known temperatures and ages form a very non-uniform sample, as
they were discovered by different methods with different instruments.
Different selection effects are in the game.

D). Mostly,  objects with known age and temperature are younger than $10^5$ years.

E). There are some additional pieces of data which are not used in the
analysis (mass distribution, etc.).

Below I briefly discuss several additional methods which can
help to improve the situation with confronting theory and observations.

The first (and the main) additional test is based on the Log N - Log S diagram.
This diagram is a useful instrument in astrophysics.
Here N represents the number of observed sources with observed 
fluxes (at some energy range) larger than
S. So, this is an integral distribution, i.e. it always grows towards
lower fluxes. 
In \cite{p2006} we proposed to use the Log N -- Log S test as an additional tool to probe theoretical cooling curves. The idea of the test is to compare the observed Log N  -- Log S distribution with the
calculated in the framework of population synthesis approach \cite{pp2004},
and to derive from this comparison if the model fits data.

Our reasoning in favour of the new test is the following:

1). Thanks to the observations made onboard the ROSAT X-ray satellite we
have a uniform sample of NSs with detected thermal radiation.

2). The test doesn't require the knowledge of ages, temperature, etc.
    Only fluxes (which are well determined) and numbers are necessary to use
this test.

3). Test is sensitive to older ($\sim 1$~Myr) sources.

4). All ingredients of the population synthesis scenario except the cooling
curves can be relatively well fixed.

One of 
the main disadvantage of the test is that one needs to have a computer code 
to test a set of cooling curves. 
The way out can be to develope a web-site
where everyone can download cooling curves and obtain the Log N -- Log S
distribution for selected parameters of the scenario. We hope to do such a
resource in future.
Another disadvantage is related to  
precision of a population synthesis model.
Not all ingredients are equaly well known, and a big piece of astrophysical
work has to be done to produce a good model. However, we believe that for 
young objects in the
solar vicinity this problem can be solved \cite{pos2006}.

Two important additions to these tests are so-called {\it brightness constraint} \cite{g2006} and {\it mass constraint} \cite{pgb2006}.
In \cite{g2006} it was proposed to take into account the fact
that despite many observational efforts very hot NSs with ages $\sim
10^3$~--~$10^4$~yrs have not been discovered. If they exist in the Galaxy, then 
it is very easy to find them (unless the interstellar absorption prevents us
to see a source, but absorption is not equaly important in all directions:
so there are relatively wide ``windows'' to observe a significant part of
the Galaxy). 
If we do not see any very hot NSs, then we have to conclude
that at least their fraction is very small. 
Indeed, we can put limits on models of
the most slowly cooling NSs. 
This means, that any model
pretending to be realistic should not produce NSs with typical masses with
temperatures higher than the observed ones. 
So, this technique is a
useful addition to the standard temperature vs. age test.

 This constraint is very sensitive to the properties of the crust of a NS
(see \cite{g2006} for details). Fitting the crust one can usualy find a
solution to satisfy the brightness constraint. 
On the other hand, it is important to remember that the Log N -- Log S test is not very
sensitive to the crust properties. The usage of only T~--~t test plus the
brightness constraint approach can lead to a wrong solution as both are not
very sensitive to the behaviour of the cooling curves for ages larger $\sim$
few $\times 10^5$ years, and just fitting the crust can help to find a
solution which can be shown to be wrong based on the Log N -- Log S test as
properties of the internal parts of a NS are not properly selected. 
We conclude, that anyway the Log N -- Log S test should be used, too, as
such complex approach helps to made a more complete testing of cooling
curves.

Mass constraint can be done if the mass spectrum of NSs is known.
Mass spectrum of NSs is an important ingredient of the population synthesis
scenario. Normally, if we consider masses in the range $1\,M_\odot < M <
2\,M_\odot$, lighter stars cool slower. 
Our estimates of the mass spectrum \cite{p2006,pgb2006} show
that the fraction of newborn NSs with masses larger than 
$\sim 1.5\, M_\odot$ is very small. This means that more massive objects
should not be used to explain observations, especially if we speak about
bright or/and typical sources. In particular, close-by young NSs, like Vela,
should not be explained as massive stars as this is very unprobable that we
are so ``lucky'' to have such a young object (age $\sim$ 10 000 years) so close.
As we show in \cite{pgb2006} this simple constraint helps to reject some
models which can successfully pass T-t or/and Log N -- Log S tests.

\section{Conclusions}

 The main conclusion is that NSs seem to appear in more flavours than it was possible to imagine even after the discovery of radio and X-ray pulsars. The Crab pulsar is not any more the most typical young NS
as the total birth rate of other types of NSs is higher than the radio pulsars birth rate.

 The main unsolved questions are related to the origin of differences between differents beasts in the Zoo of young NSs, and to possible links between them. 

 Observations of cooling NSs can help to better understand physical processes happening in superdense matter inside compact objects. New tests and constraints, hopefully, will help to succeed in selecting the actual equation of state of NSs and figure out the exact cooling mechanisms working in NSs. For details on these subjects I refer to other contributions in this volume.  

\bigskip

\noindent
{\bf Acknowledgments}
I thank my co-authors in isolated NS studies -- D. Blaschke, H. Grigorian, B. Posselt, M. Prokhorov, R. Turolla and others -- 
for fruitful collaboration and numerous discussions.
The work was supported by the RFBR grant 06-02-16025, and by the ``Dynasty'' 
and  Cariplo (Centro Volta -- Landau Network) foundations. I also thank the Organizers for partial support and hospitality.

\end{document}